\documentclass[a4paper,fleqn,usenatbib]{mnras}
\usepackage{amssymb,amsmath,graphicx,psfrag,mathtools}
\usepackage[T1]{fontenc} 
\usepackage{ae,aecompl} 
\usepackage{url}
\title[The geometry and environment of repeating FRBs]{On the geometry and environment of repeating FRBs}
\author[Du et al.]{Shuang Du$^{1,2}$\thanks{E-mail: dushuang@pku.edu.cn}, Weihua Wang$^{1,2}$, Xuhao Wu$^{1,2}$, Renxin Xu$^{1,2}$\thanks{E-mail:r.x.xu@pku.edu.cn}\\
$^{1}${State Key Laboratory of Nuclear Physics and Technology, School of Physics, Peking University, Beijing 100871, China}\\
$^{2}${Kavli Institute for Astronomy and Astrophysics, Peking University, Beijing 100871, China}}
\date{\today}
\begin{document} 
\label{firstpage} 
\pagerange{\pageref{firstpage}--\pageref{lastpage}} 
\maketitle 
\begin{abstract}
We propose a geometrical explanation for periodically and nonperiodically repeating fast radio bursts (FRBs) under neutron star (NS)-companion systems.
We suggest a constant critical binary separation, $r_{\rm c}$, within which the interaction between the NS and companion can trigger FRB bursts.
For an elliptic orbit with the minimum and maximum binary separations, $r_{\rm min}$ and $r_{\rm max}$,
a periodically repeating FRB with an active period could be reproduced if $r_{\rm min}<r_{\rm c}<r_{\rm max}$.
However, if $r_{\rm max}<r_{\rm c}$, the modulation of orbital motion will not work due to persistent interaction,
and this kind of repeating FRBs should be nonperiodic.
We test relevant NS-companion binary scenarios on the basis of FRB 180916.J0158+65 and FRB 121102 under this geometrical frame.
It is found that the pulsar-asteroid belt impact model is more suitable to explain these two FRBs since this model is compatible with different companions (e.g., massive stars and black holes).
At last, we point out that FRB 121102-like samples are potential objects which can reveal the evolution of star-forming region.
\end{abstract}
\section{Introduction}
The origin of fast radio bursts (FRBs, \citealt{2007Sci...318..777L,2013Sci...341...53T} ) is still mysterious (see \citealt{2018PrPNP.103....1K} and \citealt{2019A&ARv..27....4P} for reviews)
but observations continue to refresh the understanding of FRBs.
For example, the recent detection of a $\sim 16\;\rm day$ period from FRB 180916.J0158+65
indicates the progenitor of this FRB may be either a neutron star (NS)-companion binary or a precessing NS \citep{CHIME2020,2020arXiv200402862C}
since the size of a non-relativistically moving source should be smaller than$\sim 10^{7}\;\rm cm$ as evident from FRB durations\footnote{
In principle, black hole binaries (the present model of black hole binaries applies to one-off FRBs; see, \citealt{2016ApJ...827L..31Z}) and accreting black holes with precessing jets \citep{Katz2020}
could also provide the small scale radiation regions and periods of periodically repeating FRBs.
At present, no certain mechanism and observation shows a strong coherent and
millisecond-duration radio pulse can be emitted from these systems.}.
The follow-up observation of the previous nonperiodically repeating FRB 121102 \citep{2016Natur.531..202S} shows that this FRB should also be a periodically repeating FRB \citep{2020arXiv200303596R,CSS2020}.
These two observations bring up a question that are all repeating FRBs (even all FRBs) periodically repeating ones?
Before this problem is understood, we still treat FRBs as three types: one-off bursts, nonperiodically repeating bursts, and periodically repeating bursts.

Based on the different repeatability of FRBs, many progenitor models involving NSs have been proposed, e.g.,
\begin{itemize}
\item[(1)] for one-off bursts: binary NS mergers \citep{2013PASJ...65L..12T}, collapsing NSs \citep{2014AA....295..807F}
and asteroids/comets colliding with NSs \citep{2015ApJ...809...24G};
\item[(2)] for nonperiodically repeating bursts: magnetar hyper flares \citep{2014MNRAS.442L...9L},
close NS-white dwarf binaries \citep{2016ApJ...823L..28G} and
NSs ``combed" by plasma streams \citep{2017ApJ...836L..32Z};
\item[(3)] for periodically repeating bursts: asteroid belts colliding with NSs \citep{2016ApJ...829...27D}
NSs in tight O/B-star binaries \citep{2020ApJL..893L..39L}, orbital-induced precessing NSs \citep{2020arXiv200202553Y},
free/radiative precessing NSs \citep{2020ApJ...892L..15Z}, and precessing flaring magnetars \citep{2020arXiv200204595L}.
\end{itemize}
Besides, some of the models (e.g., \citealt{2016ApJ...823L..28G,2016ApJ...829...27D,2017ApJ...836L..32Z}) used to explain the early observation of FRB 121102 \citep{2016Natur.531..202S}
have been revised to reproduce the periodicity detected in FRB 180916.J0158+65 (e.g., \citealt{2020ApJ...893L..26I,2020arXiv200210478G,2020arXiv200304644D}\footnote{
\cite{2020arXiv200304644D} have already constrained the structure of the NS-asteroid belt system according to the period of FRB 180916.J0158+65. }).
These models usually focus on the observations that are related to FRB bursts themselves (e.g., duration, luminosity and period) and do not consider the observations which may reveal
the environment of FRBs (e.g., the changed/unchanged rotation measure (RM), \citealt{2018Natur.553..182M,2018PrPNP.103....1K,2019A&ARv..27....4P}).

Inspired by the consensus that long and short gamma-ray bursts are produced by similar compact star-accretion disc systems which originate from different progenitors (massive stars and NS binaries),
in this paper, we propose a general geometrical frame of NS-companion systems to explain both periodically and nonperiodically repeating FRBs
without considering a detailed radiation mechanism.
Then we will study the implications of this framework on relevant FRB models.

The remainder of this paper is organized as follows.
The details of our model (the geometry, kinematics and effect of orbital motion) are illustrated in Section \ref{sec2}.
The case studies of FRB 180916.J0158+65 and FRB 121102 are shown in Section \ref{sec3}.
We discuss the results of the two case studies in Section \ref{s4}.
Summary is presented in Section \ref{s5}.
\section{The geometrical model}\label{sec2}
\subsection{Orbital geometry}
\begin{figure}
    \centering
    \includegraphics[width=0.4\textwidth]{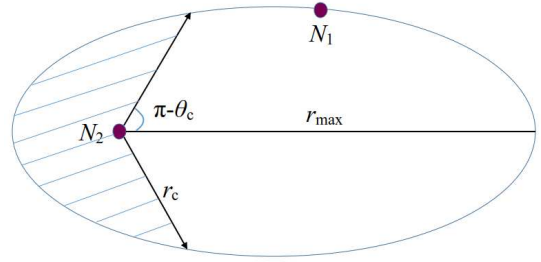}
    \caption{Schematic diagram of the geometry of our model. $N_{1}$ is the NS and $N_{2}$ is the companion.
    $r_{\rm c}$ is the constant critical binary separation (corresponding to the polar angle $\theta_{\rm c}$) under which the interaction between the NS and companion can trigger FRBs.
    $r_{\rm max}$ is the maximum binary separation.
    When the NS $N_{1}$ moves into the shaded region on the left,
     the binary begin to interact, so that FRBs are produced. }
    \label{fig1}
 \end{figure}

For an NS-companion binary with an orbital period $T$ (see Figure \ref{fig1}),
we assume there is an approximately constant critical binary separation $r_{\rm c}$ (corresponding to the polar angle $\theta_{\rm c}$)
under which the interaction between the NS $N_{1}$ and its companion $N_{2}$ can trigger bursts of a repeating FRB.
When $r_{c} < r < r_{\rm max}$ with $r_{\rm max}$ being the maximum separation between the binaries,
there is no interaction between the orbiting objects and the FRB is in quiescence.
For $r<r_{\rm c}<r_{\rm max}$, interaction between the companions will give rise to FRBs until $r$ becomes greater than $r_{\rm c}$.
For $r_{\rm c} > r_{\rm max}$, interaction will be persistent and will result in FRBs that are emitted throughout the orbit, resulting in non-periodic repeating FRBs\footnote{
If the trajectory is a parabola, this nonperiodically repeating FRB should be an ``one-off'' repeating FRB.}.


So far, no periodicity has been detected between successive pulses of repeating FRBs.
This indicates the radio emission of FRBs is not a `` lighthouse''.
The radiation should not come from the NS polar cap but from a position in the NS magnetosphere that can always be seen by the observer.
Correspondingly, the repeatability of a periodically repeating FRB during one orbital period should be mainly determined by the activity of the companion under this geometric frame.
Note that the repeatability of nonperiodically repeating FRBs does not depend on other geometric conditions as long as the condition $r_{\rm max}<r_{\rm c}$ is satisfied.
We will only discuss the geometric details of periodically repeating FRBs in the next subsection.

\subsection{Kinematics}
According to Kepler's Second Law, an elliptic orbit can be described by
\begin{eqnarray}\label{4}
r=\frac{4\pi^{2}m_{u}a^{2}b^{2}}{\alpha T^{2}}\cdot \frac{1}{1+\varepsilon\cos\theta},
\end{eqnarray}
\begin{eqnarray}\label{7}
a=\left ( \frac{\alpha T^{2}}{4\pi^{2}m_{\mu}} \right )^{1/3},
\end{eqnarray}
and
\begin{eqnarray}\label{8}
b=a\sqrt{1-\varepsilon^{2}},
\end{eqnarray}
where $m_{\mu}$ is the reduced mass of the binary (i.e., $m_{1}m_{2}/(m_{1}+m_{2})$ with $m_{1}$ being the mass of the NS and $m_{2}$ being the mass of the companion),
$a$ is the semi-major axis, $b$ is the semi-minor axis, $\alpha$ is defined as $Gm_{1}m_{2}$ with $G$ the gravitation constant,
$\theta$ is the polar angle, and $\varepsilon$ is the orbital eccentricity.

In this geometrical framework, for a repeating FRB to have a period $T$ with an active window of $\Delta T$, there should be
\begin{eqnarray}\label{5}
\frac{\Delta T}{T}\pi ab=\int_{0}^{\theta_{\rm c}}r^{2}d\theta.
\end{eqnarray}
Integrating equation (\ref{5}) gives
\begin{eqnarray}\label{6}
\frac{\Delta T}{T}\pi ab&=&\left ( \frac{4\pi^{2}m_{\mu}a^{2}b^{2}}{\alpha T^{2}} \right )^{2}\left [ \frac{A+B}{AB\sqrt{AB}}\arctan(\sqrt{\frac{B}{A}}\tan\frac{\theta}{2})\right.\nonumber\\
&&\left.-\frac{(A-B)\tan\frac{\theta}{2}}{AB\tan^{2}\frac{\theta}{2}+A^{2}}\right ]\bigg |_{0}^{\theta_{c}}
\end{eqnarray}
with $A=1+\varepsilon$ and $B=1-\varepsilon$.

Note that $T$ and $\Delta T$ are observable quantities,
if $m_{1}$, $m_{2}$ and $\varepsilon$ are known quantities
then one can solve $\theta_{c}$,
as well as the correspondingly critical binary separation $r_{\rm c}$, numerically through equation (\ref{6}).

 \begin{figure}
    \centering
    \includegraphics[width=0.45\textwidth]{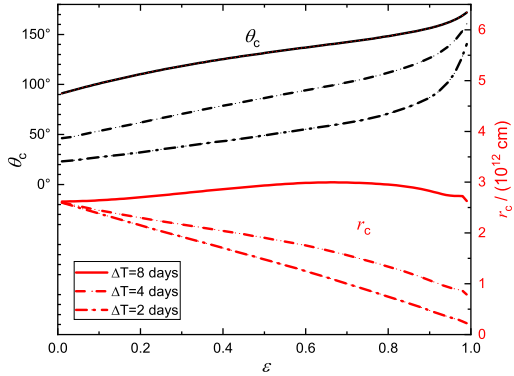}
    \caption{The values of $\theta_{c}$ and $r_{\rm c}$ versus $\varepsilon$ under $T=16\;\rm day$. The top three lines are $\theta_{c}$
    versus $\varepsilon$. The three lines at the bottom are $r_{\rm c}$ versus $\varepsilon$.
    The solid lines, dashed lines, dot-dashed lines are for $\Delta T/T=1/2$, $\Delta T/T=1/4$, $\Delta T/T=1/8$, respectively. {Note that, from equations (\ref{4}),  (\ref{7}) and  (\ref{8}),
    there is a relation that $r_{\rm c}\propto (m_{1}+m_{2})^{1/3}T^{2/3}$. Therefore, one can estimate $r_{\rm c}$ for a given binary through this relation and Figure \ref{fig2}. } }
    \label{fig2}
 \end{figure}

\subsection{The effect of the orbital motion}\label{2.3}
The orbital motion will change the binary separation, as well as the distance from the binary to the earth.
Therefore, by definition, the RM and dispersion measure (DM) could be time-varying.
For clarity, one can separate the contribution of orbital motion to the total RM and DM from observations, i.e.,
\begin{eqnarray}\label{eqx1}
{\rm RM}=\left ( \frac{{e}^{3}}{2\pi m_{\rm {e}}^{2}c^{4}} \right )\left ( \int_{0}^{l_{\rm c}}n_{\rm {e}}B_{\parallel}dl\;+ \int_{l_{\rm c}}^{d}{n_{\rm {e}}}'{B_{\parallel}}'d{l}'\right )
\end{eqnarray}
and
\begin{eqnarray}\label{eqx2}
{\rm DM}= \int_{0}^{l_{\rm c}}n_{\rm {e}}dl\;+ \int_{l_{\rm c}}^{d}{n_{\rm {e}}}'d{l}',
\end{eqnarray}
where ${e}$ and $m_{\rm {e}}$ are the charge and electron mass, $B_{\parallel}$ is the magnetic field along the line of sight,
$d$ is the shortest distance between the earth and the point in the orbit,
and $l_{\rm c}$ is the orbital-motion-induced change in distance $d$.
Since there is an inclination angle $\iota$ between the normal of the orbit and the line of sight,
one has $2b\sin \iota \leq l_{c} \leq 2a\sin \iota$.
From equations (\ref{eqx1}) and (\ref{eqx2}), the change in RM is
\begin{eqnarray}\label{eqx3}
\rm {\Delta RM}&=&\left ( \frac{{e}^{3}}{2\pi m_{\rm {e}}^{2}c^{4}} \right ) \int_{0}^{l_{\rm c}}n_{\rm {e}}B_{\parallel}dl,
\end{eqnarray}
By the definition that ${\rm {\Delta DM}}= \int_{0}^{l_{\rm c}}n_{\rm {e}}dl$
and $\bar{B}_{\parallel} {\Delta \rm DM}=\int_{0}^{l_{\rm c}}n_{\rm {e}}B_{\parallel}dl$,
equation (\ref{eqx3}) can be rewritten as
\begin{eqnarray}\label{eqx3'}
\rm {\Delta RM}=\left ( \frac{{e}^{3}\bar{B}_{\parallel}} {2\pi m_{\rm {e}}^{2}c^{4}} \right ){\rm \Delta DM}.
\end{eqnarray}
Equation (\ref{eqx3'}) predicts that
if $l_{\rm c}$ is large enough, $\Delta \rm RM$ and $\Delta \rm DM$ may change evidently during an orbital period as long as $n_{\rm {e}}$ and $B_{\parallel}$ are non-negligible

Empirically, one can adopt $B_{\parallel}=B_{\parallel,0}(l/l_{0})^{-p}$ and $n_{\rm {e}}=n_{\rm {e},0}(l/l_{0})^{-q}$ with $l_{0}$ being the size of the source which provides the magnetic environment.
Given a point source, there should be $l_{0}\ll l_{\rm c}$.  According to equation (\ref{eqx1}) and (\ref{eqx2}), one approximately has
\begin{eqnarray}\label{eqx4}
&&\left ( \frac{\bar{B}_{\parallel}n_{\rm {e},0}{e}^{3}} {2\pi m_{\rm {e}}^{2}c^{4}}\right )\int_{l_{0}}^{l_{\rm c}}(l/l_{0})^{-q}dl\nonumber\\
\approx&&\left ( \frac{B_{\parallel,0}n_{\rm {e},0}{e}^{3}}{2\pi m_{\rm {e}}^{2}c^{4}} \right ) \int_{l_{0}}^{l_{\rm c}}(l/l_{0})^{-p-q}dl.
\end{eqnarray}
Integrating equation (\ref{eqx4}) gives
\begin{eqnarray}\label{eqx5}
B_{\parallel,0}\approx\bar{B}_{\parallel}\cdot\frac{p+q-1}{1-q}\cdot \left ( \frac{l_{\rm c}}{l_{0}} \right )^{-q+1}
\end{eqnarray}
under $0<q<1$, $-p-q+1<0$, and
\begin{eqnarray}\label{eqx6}
B_{\parallel,0}\approx\bar{B}_{\parallel}\cdot\frac{p+q-1}{q-1}
\end{eqnarray}
under $q>1$, $-p-q+1<0$.

Equations (\ref{eqx3'}) and (\ref{eqx4}) indicate that the changes in RM and DM may provide the information of magnetic environment of periodically repeating FRBs.
In the next, we will study two FRB samples on the basis of the above discussion.

\section{Two case studies}\label{sec3}
\subsection{FRB 180916.J0158+65}\label{3.1}
FRB 180916.J0158+65 shows a period of $T\approx 16\;\rm day$ and an active period of $\Delta T\approx 4\;\rm day$ \citep{CHIME2020}. According to equations (\ref{4}) and (\ref{6}),
the values of $\theta_{c}$ and $r_{\rm c}$ versus $e$ are shown in Figure \ref{fig2} (see Appendix A for more details)
by assuming the companion also is an NS and $m_{1}=m_{2}=1.4\;\rm M_{\odot}$.
Since the companion may also be a massive star/black hole/white dwarf,
we discuss these scenarios separately.

Under this geometrical frame, the interaction between the NS and companion (e.g., the accretion/wind interaction)
should provide an approximately constant critical separation $r_{\rm c}$.
For the NS-massive star binary scenario, the accretion/wind interaction is susceptible to
the activity of the massive star, so that the critical separation $r_{\rm c}$ may not be constant with time.
The spin-down power of an NS is nearly a constant, as well as the critical separation induced by this wind interaction.
Therefore, as the companion, an NS is worthy of consideration.

Under NS-NS scenario, the wind from the companion NS $N_{2}$ should be strong enough to ``comb" the NS $N_{1}$ \citep{2020ApJ...890L..24Z}, i.e.,
\begin{eqnarray}\label{9}
\frac{L_{\rm sd,2}}{4\pi r_{\rm 2}^{2}c}=\frac{B_{\rm p,1}^{2}}{8\pi}\left ( \frac{R_{\ast}}{r_{1}} \right )^{6},
\end{eqnarray}
where $r_{1}$ and $r_{2}$ are distances of the interaction front from NSs $N_{1}$ and $N_{2}$, respectively,
$L_{\rm sd,2}$ is the spin-down power of NS $N_{2}$, $B_{\rm p,1}$ and $R_{\ast}$ are the polar cap magnetic field and radius of NS $N_{1}$,
and $c$ is the speed of light.
The typical isotropic value of repeating FRBs, $E_{\rm FRB,iso}$, is a few $10^{41}\;\rm erg\; s^{-1}$  \citep{LLKM2020}.
In principle, the rotational energy of the NS can satisfy this energy requirement
but there is no clear mechanism to extract this energy through such an interaction between the two NSs.
We turn to consider magnetic energy (which can be dissipated through magnetic reconnection).
The magnetic-energy density of NS $N_{1}$ at $r_{1}$ should be high enough, i.e.,
\begin{eqnarray}\label{11}
(c\Delta t)^{3}\frac{B_{\rm p,1}^{2}}{8\pi}\left ( \frac{R_{\ast}}{r_{1}} \right )^{6}\sim f_{\rm b}E_{\rm FRB,iso},
\end{eqnarray}
where $\Delta t$ is the duration of a FRB burst, and $f_{\rm b}$ is the beaming factor.
Equation (\ref{11}) gives
\begin{eqnarray}\label{12}
B_{\rm p,1}\sim 3\times 10^{12}\left ( \frac{f_{\rm b}E_{\rm FRB,iso}}{10^{40}\;\rm erg} \right )
\left ( \frac{\Delta t}{1\;\rm ms} \right )^{-3}\left ( \frac{r_{1}}{10^{7}\;\rm cm} \right )\;\rm G,
\end{eqnarray}
where $R_{\ast}=10^{6}\;\rm cm$ is adopted for the estimation.

According to equation (\ref{12}), $r_{1}$ should be much smaller than $r_{\rm c}$ (e.g., $r_{1}\sim c\Delta t$),
otherwise $B_{\rm p,1}$ will be too strong.
Correspondingly, $r_{2}$ is given by
\begin{eqnarray}
r_{2}=r_{\rm c}-r_{1}\sim r_{\rm c}-c\Delta t \sim r_{\rm c}\sim  10^{12}\;\rm cm.
\end{eqnarray}
However, this brings up another problem that
the magnetic field of NS $N_{2}$ would be unreasonable unless there is a very small $f_{\rm b}$\footnote{It is unrealistic since the size of the wind from the companion should be larger than the radius of NS $N_{1}$. Half of the magnetosphere of NS $N_{1}$ should be disturbed.},
because, according to equation (\ref{9}), there is
\begin{eqnarray}\label{13}
L_{\rm sd,2}&\sim& 4\pi r_{2}^{2}c\cdot \frac{f_{\rm b}E_{\rm FRB,iso}}{c^{3}\Delta t^{3}}\nonumber\\
&\sim& 10^{53}\left ( \frac{f_{\rm b}E_{\rm FRB,iso}}{10^{40}\;\rm erg} \right )
\left ( \frac{\Delta t}{1\;\rm ms} \right )^{-3}\;\rm erg\;s^{-1}.
\end{eqnarray}
Equation (\ref{13}) shows the NS $N_{2}$ must be a millisecond magnetar with $B_{\rm p,1}\sim 10^{16}\;\rm G$.
This powerful wind only can last for $<1 \;\rm s$ since the largest rotational energy of an NS is $\sim 10^{52}\;\rm erg$.

For the NS-black hole binary scenario, both the accretion interaction and wind interaction require an accreting black hole.
However, the changes in the accretion disc can result in changes in $r_{\rm c}$.
Besides, the luminosity of a super-Eddington accreting black hole is much smaller than that of a millisecond magnetar (see equation \ref{13}),
so the wind from this accreting black hole is not strong enough to perturb the magnetosphere of NS $N_{1}$.

For the NS-white dwarf scenario, the wind from the white dwarf is much weaker.
FRB bursts should be triggered by the accretion interaction.
Since white dwarfs do not have mass ejections and bursts as the sun,
the accretion interaction between the NS and white dwarf could be different from
that of the NS-massive star scenario. The separation $r_{\rm c}$ under this case may be approximately a constant.
However, for the specific NS-white dwarf binary model \citep{2020arXiv200210478G}, an extremely high eccentricity $(\varepsilon > 0.95)$ is required to explain FRB 180916.J0158+65.
This model demands the FRBs with $T> 16\;\rm day$ to be special ones. Therefore, it can be tested after enough periodically repeating FRBs are detected.

In summary, the above discussion shows that the NS-massive star binary scenario can not provide a constant $r_{\rm c}$;
the NS-NS binary scenario and NS-black hole binary scenario can not provide strong winds;
and the NS-white dwarf binary model \citep{2020arXiv200210478G} requires a special orbit with extremely high eccentricity.
These models cannot explain all the observed characteristics of FRBs.

However, it is worth noting that the accretion/wind interaction is not the only way to provide a critical separation
as long as the above stellar-mass objects have asteroid belts.
Under the pulsar-asteroid belt impact model \citep{2016ApJ...829...27D,2020arXiv200304644D},
the outer boundary of the asteroid belt is naturally corresponding to the critical separation $r_{\rm c}$.
Besides, there is another critical radius, $r_{\rm c}^{\prime}$ , i.e., the inner radius of the asteroid belt.
If the trajectory of the NS can cross the inner boundary, the asteroid belt will divide the binary separation into three segments,
i.e., $r<r_{\rm c}^{\prime}$, $r_{\rm c}^{\prime}<r<r_{\rm c}$ and $r_{\rm c}<r$.
Therefore, there will be two periodic active phases which are separated by a quiescent phase during one orbital period.
We suggest to fold periodically repeating FRBs at their period just like that of \cite{CHIME2020}.
If such a FRB is found, the other FRB models should at least complement the corresponding details (e.g., for precessing-NS models,
the precession angle of the FRB beam is larger than the opening angle of the FRB beam).
On the other hand, there is a tiny probability that the orbit of the NS $N_{1}$ happens to be in the asteroid belt (corresponding to $r_{\rm max}<r_{\rm c}$),
so that the induced repeating FRB will show aperiodicity.
Hence, the NS-astroid-belt model predicts that the number of aperiodically repeating FRBs will be much less than the periodically repeating ones.

So far, no observation shows that the RM and DM of FRB 180916.J0158+65 have obvious evolution.
If the RM and DM of FRB 180916.J0158+65 are almost constants,
according to equation (\ref{eqx3}), there are two explanations: (a) $l_{\rm c}$ is small enough; (b) $n_{\rm e}$ and $B_{\parallel}$ are negligible.
Given $2b \sin \iota \leq l_{c}\leq 2a\sin \iota$, $l_{\rm c}$ can only be neglected when $\iota$ is very small (the orbit happens to be face-on).
So, $l_{c}$ is more likely a non-negligible quantity.
The explanation (b) should be more reasonable, i.e., the companion is at least weak magnetized (e.g., a massive star/ black hole).
Alternatively, if future follow-up observation confirms this unchanged RM and DM, the single precessing NS scenario
(corresponding to explanation (a); see, e.g., \citealt{2020ApJ...892L..15Z,2020arXiv200204595L}) is more suitable for explaining this observation.

\subsection{FRB 121102}\label{3.2}
FRB 121102 is the first localized FRB \citep{2017ApJ...834....L7}. The long-time follow-up observation shows that
FRB 121102 also is a periodically repeating FRB with $T\sim 160\;\rm day$ and $\Delta T\sim 76\;\rm day$ \citep{2020arXiv200303596R,CSS2020}.
Note that $r_{\rm c}\propto (m_{1}+m_{2})^{1/3}T^{2/3}$ and $r_{\rm c}$ is not sensitive to $\varepsilon$ (see Figure \ref{fig2}) when $\Delta T/T\sim 1/2$.
For an NS binary scenario, there is $r_{\rm c}\sim 7\times 10^{13}\;\rm cm$.
Comparing with the case of FRB 180916.J0158+65, this time the accretion/wind interaction must be stronger since $l_{\rm c}$ gets longer.
Therefore, under the wind interaction, the NS-NS binary and NS-black hole binary scenarios are more powerless to explain FRB 121102 according to the discussion in Section \ref{3.1}.
Under the accretion interaction, the interaction between the NS and white dwarf should work on a longer distance.
Besides, the longer period of FRB 121102 indicates a much larger eccentricity and a much smaller white dwarf for the certain NS-white dwarf binary model
(this is unreasonable; see Figure 2 of \citealt{2020arXiv200210478G}).
Comparing with the above scenarios and models, the pulsar-asteroid belt impact model seems to be not quite that extreme
(it needs a huge asteroid belt; see \citealt{2019MNRAS.485.1367S}).

Observations have shown that the RM of FRB 121102 changed from $1.46\times 10^{5}\;\rm rad\cdot m^{-2}$
to $1.33\times 10^{5}\;\rm rad\cdot m^{-2}$ within $7 $ months \citep{2018Natur.553..182M}.
The recent work shows that the RM of FRB 121102 has a consistent decreasing trend in RM with the DM being steadily increasing \citep{2020arXiv200912135H}.
Thus, despite the orbital motion could induce the change in RM, this effect should not be the primary cause since the orbital period is much shorter than the duration of the decrease in RM.
Nevertheless, we can use the published data \citep{2018Natur.553..182M} to estimate the upper limit, $B_{\rm \parallel,0,max}$, of $B_{\parallel,0}$.
We will roughly adopt $l_{\rm c}\sim r_{\rm c}$ (see the last paragraph of Section \ref{3.1}) for the following estimation.

Away from a point source, the radial component of the magnetic field decays as $l^{-2}$,
and the toroidal component decays as $l^{-1}$, i.e., $1<p<2$ (see, e.g., \citealt{SDD2001}).
On the other hand, $q$ should be $\sim 0$ for intergalactic medium and $\sim 2$ for stellar wind.
In any case, the middle term of the right side of equation (\ref{eqx5}) is larger than $1$.
Therefore, $\bar{B}_{\parallel}$ should be small enough to keep the value of $B_{\parallel,0}$ reasonable (see equation \ref{eqx5}).
From equation (\ref{eqx3'}), there is (see also \citealt{2018PrPNP.103....1K})
\begin{eqnarray}\label{20}
\bar{B}_{\parallel}&=&\left ( \frac{2\pi m_{\rm {e}}^{2}c^{4}}{e^{3}} \right )\frac{\rm {\Delta RM}}{\rm {\Delta DM}}\nonumber\\
&=&67.6\left ( \frac{\rm {\Delta RM}}{10^{4} \;\rm rad\cdot m^{-2}} \right )\left ( \frac{3 \;\rm pc\cdot cm^{-3}}{\rm {\Delta DM}} \right )\;\rm mG.
\end{eqnarray}
Since $\Delta \rm RM$ induced by the orbital motion should be smaller than $1.46\times 10^{5}\;\rm rad\cdot m^{-2}-1.33\times 10^{5}\;\rm rad\cdot m^{-2}=1.3\times 10^{4}\;\rm rad\cdot m^{-2}$,
we can estimate $B_{\rm \parallel,0,max}$ under different companions through equations (\ref{eqx5}), (\ref{eqx6}) and (\ref{20}).

Under the intergalactic medium situation, the results are as follows.
\begin{itemize}
\item[(i)] For the NS-NS binary scenario, there is $l_{0}\sim 10^{6}\;\rm cm$. So one has
\begin{eqnarray}\label{B1}
B_{\rm \parallel,0,max}\sim 5.2\times 10^{5}\left ( \frac{\rm \Delta DM}{3 \;\rm pc\; cm^{-3}} \right )^{-1}\left ( \frac{l_{\rm c}}{10^{13}\;\rm cm} \right )\;\rm G.
\end{eqnarray}
However, observations\footnote{Data comes from The ATNF Pulsar Database
(https://www.atnf.csiro.au/research/pulsar/psrcat/)} show that the magnetic field of the NS in an NS-NS binary is stronger than $10^{9}\;\rm G$.
Therefore, the magnetic field of an NS is too large for equations (\ref{B1})\footnote {Even if the value of $\rm \Delta DM$ is taken as $\sim 0.1$ (The CHIME/FRB Collaboration et al. 2020),
the value of $B_{\parallel,0}$ is still much smaller than $10^{9}\;\rm G$. }.
\item[(ii)] For the NS-massive star binary scenario, we adopt $l_{0}\sim 10^{11}\;\rm cm$. Then one has
\begin{eqnarray}\label{B2}
B_{\rm \parallel,0,max}\sim 5.2\left ( \frac{\rm \Delta DM}{3 \;\rm pc\; cm^{-3}} \right )^{-1}\left ( \frac{l_{\rm c}}{10^{13}\;\rm cm} \right )\;\rm G.
\end{eqnarray}
This value is compatible with the magnetic field of a massive star \citep{2009MNRAS..394....1338B}.
\item[(iii)] For the NS-white dwarf binary scenario, $l_{0}\sim 10^{8}\;\rm cm$ is adopted for estimation. There is
\begin{eqnarray}\label{B3}
B_{\rm \parallel,0,max}\sim 5.2\times 10^{3}\left ( \frac{\rm \Delta DM}{3 \;\rm pc\; cm^{-3}} \right )^{-1}\left ( \frac{l_{\rm c}}{10^{13}\;\rm cm} \right )\;\rm G.
\end{eqnarray}
This value also is compatible with observations \citep{2008MNRAS.387..897T}.
\item[(iiii)] If the companion is a black hole, the magnetic field should be provided by the accretion disc.
We adopt the outer boundary of the disc $\sim 50r_{\rm g}$ with $r_{\rm g}$ being the Schwarzschild radius of the black hole.
The result is
\begin{eqnarray}\label{B4}
B_{\rm \parallel,0,max}\sim 17\left ( \frac{\rm \Delta DM}{3 \;\rm pc\; cm^{-3}} \right )^{-1}\left ( \frac{l_{\rm c}}{10^{14}\;\rm cm} \right )\left ( \frac{M_{\rm BH}}{10^{3}\rm M_{\odot }} \right )^{-1}\;\rm G,
\end{eqnarray}
where $M_{\rm BH}$ is the mass of the black hole.
This value is consistent with previous work (e.g., \citealt{1995AA....295..807F}).
\end{itemize}

Under the stellar wind situation, the value of $B_{\parallel,0}$ is independent of the companion (see equation \ref{eqx6}).
Since the magnetic field of an NS or a white dwarf is too large for equation (\ref{20}), the companion should be a massive star or a black hole.

\section{Comparison of the two case studies}\label{s4}
Based on the study of FRB 180916.J0158+65 presented in the previous section, if the FRB is induced by the accretion/wind interaction,
the companion should not be a massive star or an NS;
the companion star could be a white dwarf only if FRB 180916.J0158+65 is a special one.
The unchanged RM of FRB 180916.J0158+65 indicates the companion should be weakly magnetized.
Under the pulsar-asteroid belt impact model, the companion could be a massive star/black hole as long as the companion has an asteroid belt.

The case study of FRB 121102 shows that the feasibilities of scenarios involving accretion/wind interaction (e.g., NS-NS/white dwarf binary scenario)
need some unreasonable conditions due to the larger period $T$ and critical separation $r_{\rm c}$.
The pulsar-asteroid belt impact model
could reproduce the observed $T$, $\Delta T$ and satisfy the change in RM more reasonably (the asteroid belt should be large enough) due to the compatibility with different companions,
e.g., massive stars and black holes.

In Section \ref{3.2}, we mention that orbital motion is not the primary cause to induce the change in the RM of FRB 121102.
Since the source of FRB 121102 is co-located with a star-forming region \citep{2017ApJ...843L...8B},
the gases in the star-forming region may mainly induce the change in the RM of FRB 121102\footnote{It is worth reminding that the source of FRB 180916.J0158+65
also is co-located with a star-forming region \citep{MNH}.
We should expect the correlation of locations between the star-forming region and the source of FRB 180916.J0158+65 to be different from that of FRB 121102.
Another explanation to the higher RM of FRB 121102 can be found in \citep{MBM} (the following discussion (e.g., equation {{x5}}) is still applicable).}.
Therefore, the following discussion can also be applied to the precessing NS scenario since the change in the RM
is induced by the evolution of the star-forming region and has nothing to do with the FRB source.
Let's check this idea at first.

The RM contributed by the star-forming region is given by
\begin{eqnarray}\label{x1}
{\rm RM}_{\rm g}=\left ( \frac{{e}^{3}}{2\pi m_{\rm {e}}^{2}c^{4}} \right ) \int_{0}^{l_{\rm g}}n_{\rm {e,g}}B_{\parallel,\rm g}dl
\end{eqnarray}
where $l_{\rm g}$ is the size scale of the gases along the line of sight, and $n_{\rm e,g}$ and $B_{\parallel,\rm g}$ are number density of electrons and magnetic field strength
along the line of sight in the gases, respectively.
To reproduce the observed RM, from equation (\ref{x1}), the magnetic field strength over the size scale $l_{\rm g}$ should be
\begin{eqnarray}\label{x2}
B_{\parallel,\rm g}\sim 12.3 \left ( \frac{{\rm RM}}{10^{5}\;\rm rad\cdot m^{-2}} \right )
\left ( \frac{l_{\rm g}}{100\;\rm pc} \right )^{-1}\left ( \frac{n_{\rm e,g}}{100\;\rm cm^{-3}} \right )^{-1}\;\rm uG.
\end{eqnarray}
If this magnetic field is provided by the dynamo process in the gases,
\begin{eqnarray}\label{x3}
\frac{B_{\parallel,\rm g}^{2}}{4\pi}< \frac{1}{2}n_{\rm p}m_{\rm p} v_{\rm p}^{2},
\end{eqnarray}
where ${n}_{\rm p}$ and ${v}_{\rm p}$ are the number density of protons and velocity of the gases, respectively,
and $m_{\rm p}$ is the proton mass.
According to equations (\ref{x2}) and (\ref{x3}), there is
\begin{eqnarray}\label{x4}
{n}_{\rm p}>12\left ( \frac{{B}_{\parallel,\rm g}}{12.3\;\rm uGs} \right )^{2} \left ( \frac{v_{\rm p}}{10^{6}\rm\; cm\;s^{-1}} \right )^{-2}\;\rm cm^{-3}.
\end{eqnarray}
Through equations (\ref{x2}) and (\ref{x4}), one can find that ${n}_{\rm p}$ is compatible with ${n}_{\rm e}$.
Therefore, this idea is self-consistent.

If the total RM is mainly contributed by the star-forming region,
according to equation (\ref{x1}),
the change in RM should be induced by the changes in $l_{\rm g}$, $n_{\rm e,g}$ and $B_{\parallel,\rm g}$.
This demands
\begin{eqnarray}\label{24}
\Delta{\rm DM}=\int_{0}^{l_{\rm g}(0)}n_{\rm e,g}(l,t)dl -\int_{0}^{l_{\rm g}(t)}n_{\rm e,g}(l,t)dl\ll {\rm DM},
\end{eqnarray}
\begin{eqnarray}\label{25}
\frac{d \rm RM}{dt}\approx \left ( \frac{{e}^{3}}{2\pi m_{\rm {e}}^{2}c^{4}} \right ) \frac{d}{dt}\left [ \int _{0}^{l_{\rm g}(t)}n_{\rm e,g}(l,t)B_{\parallel,\rm g}(l,t) dl\right ],
\end{eqnarray}
where $t$ is the time since the first measurement of RM.
Note that
\begin{eqnarray}
\frac{\rm \Delta DM}{\rm DM}\ll \frac {\rm \Delta RM}{\rm RM},
\end{eqnarray}
${\rm DM}_{\rm g}=\int _{0}^{l_{\rm g}(t)}n_{\rm e,g}(l,t)dl$ can be approximately treated as a constant (see equation \ref{24}).
Therefore, equation (\ref{25}) is reduced to
\begin{eqnarray}\label{x5}
\frac{d \rm RM}{dt} &\approx& \left ( \frac{{e}^{3}}{2\pi m_{\rm {e}}^{2}c^{4}} \right ) \frac{d}{dt}\left [ {\rm DM}_{\rm g}\bar{B}_{\parallel,\rm g}(t)\right ]\nonumber\\
&=&\left ( \frac{{e}^{3}}{2\pi m_{\rm {e}}^{2}c^{4}} \right )  {\rm DM}_{\rm g}\frac{d}{dt}\left [ \bar{B}_{\parallel,\rm g}(t) \right ],
\end{eqnarray}
where $\bar{B}_{\parallel,\rm g}$ has the same definition as that of $\bar{B}_{\parallel}$.
So, once the time evolution of RM is determined by future observations, one can infer the time evolution of $ \bar{B}_{\parallel,\rm g}$ through equation (\ref{x5}).
FRB121102-like samples may be potential objects that can probe the evolution of star-forming regions in distant galaxies (e.g., turbulence, convection).

\section{Summary}\label{s5}

In this paper, we show a general geometrical frame to explain the periodically and nonperiodically repeating FRBs.
We study FRB 180916.J0158+65 and FRB 121102 under this geometrical frame and find that the pulsar-asteroid belt impact model is preferred
(although a huge asteroid belt is needed, \citealt{2019MNRAS.485.1367S,2020arXiv200304644D}).
Besides, we point out that FRB 121102-like samples may be potential objects which can reveal the evolution of star-forming region.

Although we concentrate on the geometrical frame of NS-companion systems in this paper,
it is worth reminding that the precessing NS scenario is more suitable for explaining a repeating FRB with an unchanged RM.
We also discuss a possible explanation to the changed RM of FRB 121102 under the precessing NS scenario in Section \ref{s4}.
This is only one aspect of the problem. The invoking of a precessing NS is to produce a gyroscope-like radio beam so that
the beam toward/outward Earth's field of view can also reproduce the observed periodicity (e.g., \citealt{2020ApJ...892L..15Z,2020arXiv200204595L}).
However, there is no conclusive evidence that shows that precession exists in the known isolated pulsars and magnetars on such short timescales till now\footnote{
The spin-precession period induced by spin-orbit coupling is too long for periodically repeating FRBs (even for the most compact relativistic system: PSR J0737-3039 \citep{2003Natur.426..531B,2004Sci...303.1153L}).}.
Besides, the duty cycle $\Delta T/T$ depends on the size of the radio-emission region on the NS
(see, e.g., the pink semicircle in Figure 3 of \citealt{2020ApJ...892L..15Z}).
If the result of \cite{2020arXiv200303596R} is confirmed, this scenario will face a problem that the emission region becomes unrealistic for a 160-day periodicity
since $\Delta T/T$ is the ratio of the radio-emission region size to the circumference at the same latitude.
Maybe, the free/radiative precessing NS model \citep{2020ApJ...892L..15Z} needs a wider radio beam;
and the precessing flaring magnetar model needs a wider ``pancake"-like plasmoid to produce a FRB beam with a larger solid angle (see the lower panel of Figure 1 in \citealt{2020arXiv200204595L}).

Nevertheless, both the geometric frame and the model invoke a gyroscope-like radio beam
have to explain the lack of FRBs in the Milky Way (see Appendix B for more discussions).
There are three speculations for the no detection:
(i) such a system should be a special one that belongs to ``rare species'' so that the absolute quantity of these systems is much less than the number of NSs in the Milky Way;
(ii) the radio emissions of these rare-species systems tend to be outward rather than along the galactic disc so that the Galactic FRBs are difficult to be seen;
(iii) the conditions for coherent radiation are hard to be satisfied (suitable magnetic field, position and charge density, etc.) since not every X-ray burst is corresponding to a radio burst.

In this paper, we do not discuss the detailed radiation mechanism of radio emission (e.g., \citealt{2019ApJ...876...15W}) since it depends on the unknown structure of NS magnetospheres and
complicated magnetohydrodynamic processes.
Although the details of radio radiation are unknown, this NS-companion frame can still be tested by detecting the gravitational-wave radiation induced by orbital inspiral
(e.g., LISA \citep{2017arXiv170200786A}, TianQin \citep{2016CQGra..33c5010L} and Taiji \citep{2018arXiv180709495R}).

\section{acknowledgments}
We would like to thank the anonymous referee for
his/her very useful comments that have allowed us
to improve our paper (e.g., the suggestion to estimate the effect of the change in $r_{\rm c}$ on $\Delta T/T$ and corrections of English expression).
We would like to thank Mr. Weiyang Wang for telling us that
there is a 4-day active period of FRB 180916.J0158+65.
This motivates us to construct such a geometric frame.
We would like to thank Prof. Yuefang Wu and Mr. Heng Xu for useful discussion.
This work was supported by the National Key R\&D
Program of China (Grant No. 2017YFA0402602), the National
Natural Science Foundation of China (Grant Nos. 11673002, and
U1531243), and the Strategic Priority Research Program of
Chinese Academy Sciences (Grant No. XDB23010200).

\begin{appendix}
\section{}\label{ap1}
From equations (\ref{4}),  (\ref{7}) and  (\ref{8}), there is
\begin{eqnarray}\label{A1}
\frac{\partial r_{\rm c}}{\partial \varepsilon}=\left ( \frac{\alpha T^{2}}{4\pi^{2}m_{\rm u}} \right )^{1/3}\frac{\varepsilon^{2}\cos\theta_{\rm c} +2\varepsilon+1}{(1+\varepsilon\cos\theta_{\rm c} )^{2}}.
\end{eqnarray}
If changes in $r_{\rm c}$ on the size scale of an NS magnetosphere does not affect $\Delta T/T$, there should be a ``step length" of the eccentricity, $\Delta \varepsilon$, which satisfies
\begin{eqnarray}\label{A2}
\frac{\partial r_{\rm c}}{\partial \varepsilon}\Delta \varepsilon\sim \frac{2\pi c}{P_{1}},
\end{eqnarray}
where $P_{1}$ is the rotational period of the NS $N_{1}$.
Combining equations (\ref{A1}) and (\ref{A2}), when the change in orbital eccentricity is within
\begin{eqnarray}\label{A3}
\Delta \varepsilon\sim  \frac{2\pi c}{P_{1}} \left ( \frac{\alpha T^{2}}{4\pi^{2}m_{\rm u}} \right )^{-1/3}\frac{(1+\varepsilon\cos\theta_{\rm c} )^{2}}{\varepsilon^{2}\cos\theta_{\rm c} +2\varepsilon+1},
\end{eqnarray}
$\Delta T/T$ would be approximately a constant.
Considering $r_{\rm c}\sim  10^{11} - 10^{12}\;\rm cm$ and $P_{1}=10^{-2}-10^{-1}\;\rm s$, there is $\Delta \varepsilon\sim  10^{2}-10^{4}$ according to equation (\ref{A3}).
This is not a reasonable result since $\Delta \varepsilon$ must be smaller than $1$.
Under the geometrical frame, the unreasonable value of $\Delta \varepsilon$ indicates a small change in $r_{\rm c}$  can and must affect $\Delta T/T$.

\section{}\label{ap2}
We do not believe FRB 200428 \citep{2020arXiv200510828B,CHIME2020b} has the same origin as the cosmological FRBs.
Before discussing this question further, we should specify that what is an FRB.
If an FRB is just defined as a millisecond-duration bright radio pulse, it is fine to call the two radio pulses \citep{2020arXiv200510828B,CHIME2020b} as FRBs.
However, once considering the physical origin, one should treat the differences between FRB 200428 (weaker luminosity and X-ray burst association) and cosmological FRBs more carefully
although the absence of FRB 200428-like cosmological FRBs can be naturally explained as a selection effect.
Remember that soft gamma-ray repeaters are mistakenly believed to be gamma-ray bursts in history.
If all FRBs are produced by the events that generate SGR bursts \citep{2020arXiv200511071L,2020ApJ...898L..29M,2020arXiv200511178R,2020arXiv200512164T},
it is difficult to reconcile the association between accidental SGR bursts and periodically repeating FRBs (unless the period origins from a certain``external factor'', e.g., asteroid belts).
As the reviewer comments ``the fact that luminosity of the SGR burst is at least 30 times smaller than the faintest pulse of an FRB
is a strong enough argument to suggest that not all repeaters may come from SGR like origin".

\end{appendix}


\bsp 
\label{lastpage} 
\end{document}